\documentstyle{mn}
\outer\def\gtae {$\buildrel {\lower3pt\hbox{$>$}} \over 
{\lower2pt\hbox{$\sim$}} $}

\title{Simultaneous optical polarimetry and X-ray observations of 
the magnetic CV CP Tuc (AX J2315--592)}
\author[G. Ramsay, S. B. Potter, D. A. H. Buckley, P. Wheatley]
{Gavin Ramsay$^{1}$, S. B. Potter$^{1}$, David A. H. Buckley$^{2}$, 
Peter J. Wheatley$^{3}$\\
$^{1}$Mullard Space Science Laboratory, University College London,
Holmbury St.Mary, Dorking, Surrey, RH5 6NT\\
$^{2}$South African Astronomical Observatory, PO Box 9, Observatory 7935, Cape
Town, South Africa\\
$^{3}$X-ray Astronomy Group, Dept of Physics \& Astronomy, University
of Leicester, University Road, Leicester, LE1 7RH\\
}

\begin{document}

\maketitle

\begin{abstract} 

CP Tuc (AX J2315--592) shows a dip in X-rays which lasts for
approximately half the binary orbit and is deeper in soft X-rays
compared with hard X-rays. It has been proposed that this dip is due
to the accretion stream obscuring the accretion region from view. If
CP Tuc was a polar, as has been suggested, then the length of such a
dip would make it unique amongst polars since in those polars in which
a dip is seen in hard X-rays the dip lasts for only 0.1 of the
orbit. We present optical polarimetry and {\sl RXTE} observations of
CP Tuc which show circular polarisation levels of $\sim$10 per cent
and find evidence for only one photometric period. These data confirm
CP Tuc as a polar. Our modelling of the polarisation data imply that
the X-ray dip is due to the bulk of the primary accretion region being
self-eclipsed by the white dwarf. The energy dependence of the dip is
due to a combination of this self-eclipse and also the presence of an
X-ray temperature gradient over the primary accretion region.

\end{abstract}


\vspace{2.5cm}

\section{Introduction}

The X-ray source AX J2315--592 (CP Tuc) was discovered by Misaki et al
(1995) using {\sl ASCA} data. Follow up observations by Thomas \&
Reinsch (1995) identified the optical counterpart ($V\sim$17) and
suggested it was a polar (or AM Her) system. These are cataclysmic
variables - CVs - in which the accreting white dwarf has a
sufficiently strong magnetic field to lock the spin of the white dwarf
into synchronous rotation with the binary orbital period. If the
photometric variation of 89 min is attributed to its orbital period
(Misaki et al 1996), then this would place it at the shorter end of
the polar orbital period distribution.

CP Tuc shows a prominent dip which in X-rays lasts approximately half
the orbital cycle (Misaki et al 1996). Misaki et al suggested that
this dip is caused by the accretion stream far from the white dwarf
obscuring our line of sight to the bright accretion region since the
dip was deeper in soft X-rays compared to hard X-rays. In all other
polars where absorption dips are seen in hard X-rays, they are visible
for only $\sim$0.1 of a cycle. In some of the polars which show these
hard X-rays dips, a much broader dip is also seen in the EUV -- these
broader dips are thought to be due to the accretion column obscuring
the accretion region and have not been seen in hard X-rays.

One of the defining properties of polars is their high level of
polarisation. To confirm the polar nature of CP Tuc we have obtained
the first optical polarimetric data of this source. To complement
these data and to determine the nature of the dip feature we obtained
quasi-simultaneous X-ray data using the {\sl RXTE} satellite (Bradt,
Rothschild, Swank 1993).

\begin{table}
\begin{center}
\begin{tabular}{lllrr}
\hline
Date&Telescope&HJD&Duration&\\
  & & Start& &\\
\hline
\multicolumn{3}{l}{\it X-ray:}\\
\hline
1997 July 19&{\it RXTE}&648.54&19.0ksec& \\
1997 July 20&{\it RXTE}&649.56&17.5ksec& \\
1997 July 23&{\it RXTE}&652.61&21.6ksec& \\
1997 July 24&{\it RXTE}&653.68&18.7ksec& \\
1997 July 29&{\it RXTE}&659.35&19.4ksec& \\
1997 July 30&{\it RXTE}&660.35&9.4ksec &\\
1997 July 31&{\it RXTE}&661.35&9.4ksec &\\
\hline
\multicolumn{3}{l}{\it CCD Photometry:}\\
\hline
1995 Nov 24/25&SAAO 1.0m&46.30& 1h55m& $B, I$\\
1995 Nov 25/26&SAAO 1.0m&47.27& 1h40m& $B, V, R$\\
\hline
\multicolumn{3}{l}{\it Optical polarimetry:}\\
\hline
1996 Sept 15/16&SAAO 1.9m&341.54& 2h33m& WL\\
1996 Sept 16/17&SAAO 1.9m&342.56& 1h31m& WL\\
1997 July 29/30&SAAO 1.9m&659.41& 6h22m& WL\\
1997 July 30/31&SAAO 1.9m&660.41& 5h45m& WL\\
1997 July 31/1&SAAO 1.9m&661.33& 2h22m& WL\\
1997 Aug 7/8&SAAO 1.9m&667.64& 59m& WL\\
1997 Aug 10/11&SAAO 1.9m&670.63& 1h5m& WL\\
\hline
\end{tabular}
\end{center}
\caption{Observing log for CP Tuc. The HJD start time is
HJD-2450000.0. WL refers to white light observations.}
\label{obslist}
\end{table}

\section{X-ray light curves \label{ephem}}
\label{ephem}

The most prominent feature of the X-ray light curve is the deep dip
first noted in {\sl ASCA} data by Misaki et al (1996). We use this dip
to derive an precise ephemeris for CP Tuc using data from {\sl ASCA}
which is in the public archive (GIS: 0.5--12keV) (Misaki et al 1996),
{\sl SAX} (MECS: 1--10keV) and {\sl RXTE} (1--20keV) (both Wheatley,
in prep). Since the data obtained using {\sl ASCA} and {\sl SAX} are
described elsewhere we will not discuss them in any detail here.

The data taken using {\sl RXTE} have not been published elsewhere and
so are described in more detail. {\sl RXTE} was launched in Dec 1995,
its prime aim being to observe sources with maximum time resolution
and moderate energy resolution. CP Tuc was detected in the 1--20keV
energy range in observations using {\sl RXTE} between 1997 July 19 and
1997 July 31 (cf table \ref{obslist}). The total usable exposure time
was 115 ksec and was spread over 7 different epochs. The mean
background subtracted count rate over the 1--20keV energy range was
12.9 ct s$^{-1}$.

Using the {\sl ASCA}, {\sl SAX} and {\sl RXTE} data we obtained a
total of 24 estimates for the time of the center of the X-ray
dip. From these timings we obtained the following linear ephemeris:

\vspace{1mm}
T = HJD 2450024.8015(7) + 0.06183207(8) $\times$ E
\vspace{1mm}

\noindent The number in brackets refers to the error on the last
digit. This ephemeris is sufficiently precise to phase all our data to
within 0.02 cycles.  This best fit period (89.0382$\pm$0.0001 mins) is
consistent with the period derived from optical photometry by Thomas
\& Reinsch (1996) (89.041$\pm$0.004 mins) and also the period derived
by Misaki et al (1996) from {\sl ASCA} data (89.34$\pm$0.78 mins). We
use the above ephemeris throughout the paper. Assuming the 89.0382 min
period is the spin period of the white dwarf then the best estimate of
the orbital period (89.051$\pm$0.014 min Thomas \& Reinsch 1996) is
consistent with the CP Tuc being synchronised and therefore a polar.

\begin{figure}
\begin{center}
\setlength{\unitlength}{1cm}
\begin{picture}(8,15)
\put(-0.5,-1.0){\includegraphics{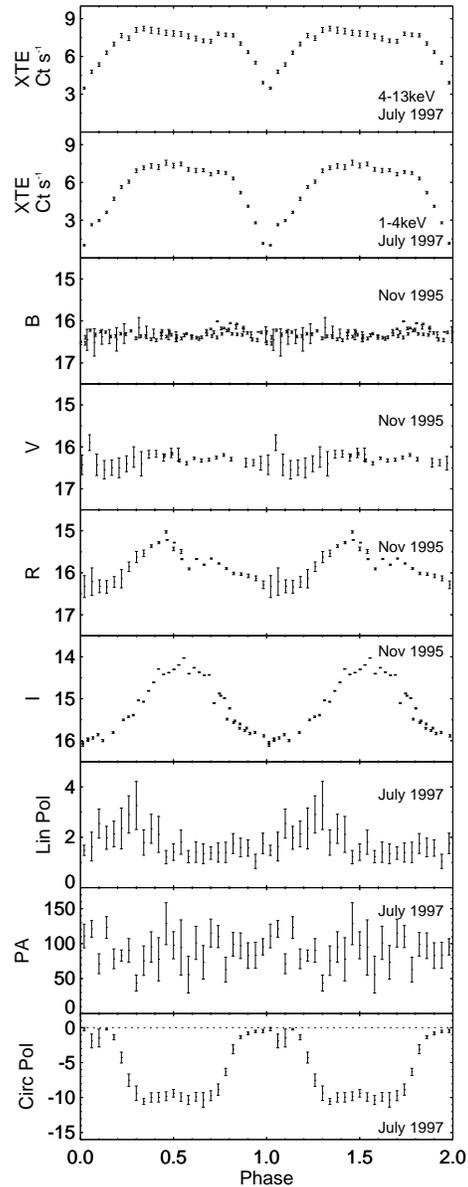}}
\end{picture}
\end{center}
\caption{ From the top -- the folded: {\sl RXTE} data (4--13keV), {\sl
RXTE} data (1--4keV), CCD photometry in $B$, $V$, $R$, and $I$ bands,
the linear polarisation, the position angle of the polarisation, and
the circular polarisation (the polarisation data were white light
observations). The X-ray and polarimetric data have been binned into
0.04 phase bins. The phasing zero corresponds to the start of the
increase in negative polarisation and is phased according to the
ephemeris in \S \ref{ephem}.}
\label{folded} 
\end{figure}

We show the background subtracted folded and binned {\sl RXTE} light
curves covering 1--4keV and 4--13keV in the upper panels of
Fig. \ref{folded}. Both light curves show a prominent dip lasting
$\Delta\phi\sim$0.45 and 0.35 cycles in the 1--4keV and 4--13keV range
respectively. The dip minimum coincides with the point in the orbital
cycle where the circular polarisation is $\sim$0\%.  The light curves
obtained using {\sl ASCA} and {\sl SAX} data have the same general
shape and features as the {\sl RXTE} light curves (not shown).

\section{Polarimetry}

\subsection{The Observations}

CP Tuc was observed at 2 epochs (cf table \ref{obslist}) using the
SAAO 1.9m telescope and UCT polarimeter (Cropper 1985). At both epochs
white light data were obtained and conditions were photometric.
Polarised and non-polarised standard stars (Hsu \& Breger 1982) and
calibration polaroids were observed at the beginning of the night to
set the position angle offsets and efficiency factors. The data were
reduced in the standard way (eg Cropper 1997).

\subsection{The folded data}

The polarisation data were folded on the spin period derived in \S
\ref{ephem} and shown in the lower panels of Fig. \ref{folded}. The
circular polarisation is at a maximum of $\sim-$10\% for around half
the orbital cycle which corresponds with the bright phase seen in the
white light intensity data (not shown). The mean linear polarisation
is 1.7$\pm$0.7 percent with a peak at $\phi\sim$0.3 of 3 percent. The
position angle is roughly constant throughout the orbital cycle.

\section{Multi-colour Photometry}

CCD photometry in the $BVR_{c}I_{c}$ system was obtained on CP Tuc
during 1995 Nov (see Table 1) using the SAAO 1.0m telescope and the
Tek8 CCD (512 x 512). Repetitive time series observations in two ($B,
I$) or three ($B, V, R$) filters were conducted, with exposure times
of 60sec. Following the usual flat-fielding and bias corrections, the
individual CCD frames were processed using the DoPHOT reduction
package (Mateo \& Schechter 1989) to derive differential magnitude
estimates. The three brightest stars on the frames were used to define
canonical frame standards, and their magnitudes derived from
photometric frames obtained on the first night (conditions were
photometric on the first night and for the first half on the second).
These magnitudes were transformed to the standard system using
observations obtained of E-region standards on the same night.  The
data were folded on the ephemeris derived in \S \ref{ephem} and are
shown in Fig. \ref{folded}. There is no significant modulation in $B$
or $V$, while in $R$ and $I$ the modulation is 1.5 mag and 2.0 mag
respectively. In $R$ and $I$ the minimum brightness occurs at the same
phase as the X-ray minimum.

\section{Photometric Variations}

\begin{figure*}
\begin{center}
\setlength{\unitlength}{1cm}
\begin{picture}(8,11.5)
\put(-5,-28){\includegraphics{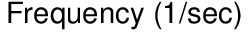}}
\end{picture}
\end{center}
\caption{From the top: the amplitude spectra of the white light
intensity data taken during July 1997 (3 nights); the data
pre-whitened using the period found by fitting the times of X-ray
minimum and its window function. The next three panels are as before
 but for the {\sl RXTE} data.}
\label{power} 
\end{figure*}

To search for all possible photometric periods we subjected the
optical photometry obtained in July 1997 to a standard Discrete
Fourier Transform. This gave an amplitude peak at 88.960 mins: the
amplitude spectrum is shown in Fig. \ref{power}.  We then pre-whitened
the data using the period found from the timings of the X-ray minimum
(89.0382 mins: \S \ref{ephem}) and its first harmonic. This
successfully removes power at the 88.960 min period indicating that
this period is consistent with the period found in \S
\ref{ephem}. Further, there is no evidence for a second period with
comparable or longer period than the 89 min spin period in the
pre-whitened spectrum. Although there is evidence for variations on
time scales shorter than the spin period in the unfolded light curve,
there is no evidence for a coherent modulation in the DFT: this
suggests these shorter period variations are due to flickering.  We
made a similar analysis using the full set of {\sl RXTE} data. We show
the amplitude spectrum and the pre-whitened data (again using the
period found in \S \ref{ephem} and its first and second harmonics) of
the {\sl RXTE} data in Fig. \ref{power}. There is no evidence for a
significant modulation at periods near or at shorter periods than the
89 min spin period. The increased power at shorter frequencies is
probably due to cycle to cycle variations. We make a similar
conclusion from an analysis of the circular polarisation data taken in
July and August 1997.

\section{Modelling the polarisation data \label{polmod}}
\label{polmod}

To model our polarisation data we fitted the data using the
optimisation method of Potter, Hakala \& Cropper (1998). Until very
recently polarisation data was modeled by constructing accretion
regions on the surface of the white dwarf by hand and then a good fit
to the data was achieved by trial and error. The method of Potter,
Hakala \& Cropper finds the best fit to the data and maps the shape,
location and structure of the cyclotron emission region(s) in an
objective manner. The best fit is the simplest solution found using a
maximum entropy technique. Strictly speaking the model maps the
accretion region(s) in terms of an optical depth parameter (a
dimensionless parameter describing density/ optical depth). In this
paper we assume a simple dipole magnetic field for the white dwarf.
While it is possible that the geometry of the magnetic field maybe
more complex, we do not consider the additional parameters that an
off-set dipole (or multi-pole model) would require are justified in
this case.

A wide range of parameter space was initially searched: the
inclination of the system, $i$, was searched between 0--80$^{\circ}$
(there is no evidence for an eclipse), the angle between the spin axis
and the magnetic axis, $\beta$, was searched between 0--90$^{\circ}$
and the phase at which the magnetic dipole crosses our line of sight
is searched through all phases. The magnetic field strength at each
pole was fixed at 15MG and the shock temperature was fixed at 17keV
(Thomas \& Reinsch 1996).

\begin{figure}
\begin{center}
\setlength{\unitlength}{1cm}
\begin{picture}(5,10)
\put(-4.5,-2){\includegraphics{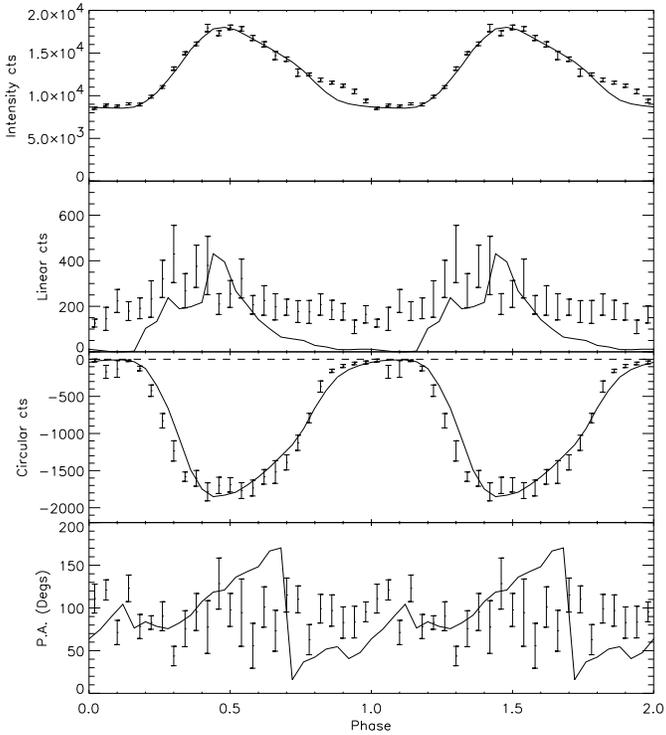}}
\end{picture}
\end{center}
\caption{The polarisation data (white light) together with the model
fit where $i=42^{\circ}$ and $\beta=50^{\circ}$.}
\label{polfit} 
\end{figure}

For each model fit to the data a goodness of fit was calculated.  We
found that we were not able to place tight constraints on the geometry
of the system: $i$ was found to be \gtae20$^{\circ}$, the phase that
the magnetic dipole crosses our line of sight clustered around
$\phi\sim$0.3 and $\beta$ was not constrained. We show in
Fig. \ref{polfit} one of the best model fits ($i$=42$^{\circ}$ and
$\beta$=50$^{\circ}$). The fits to the circular polarisation and
intensity data is reasonably good. During the bright phase the model
linear polarisation has the same general level of polarisation as the
data. During the faint phase the model underestimates the linear
polarisation. This is partly due to the fact that the linear polarised
flux is much less than the circularly polarised flux and hence the
optimisation method we use gives greater weight to the circular
polarisation curves in the fitting process.

\begin{figure*}
\begin{center}
\setlength{\unitlength}{1cm}
\begin{picture}(8,7)
\put(-5,12){\includegraphics{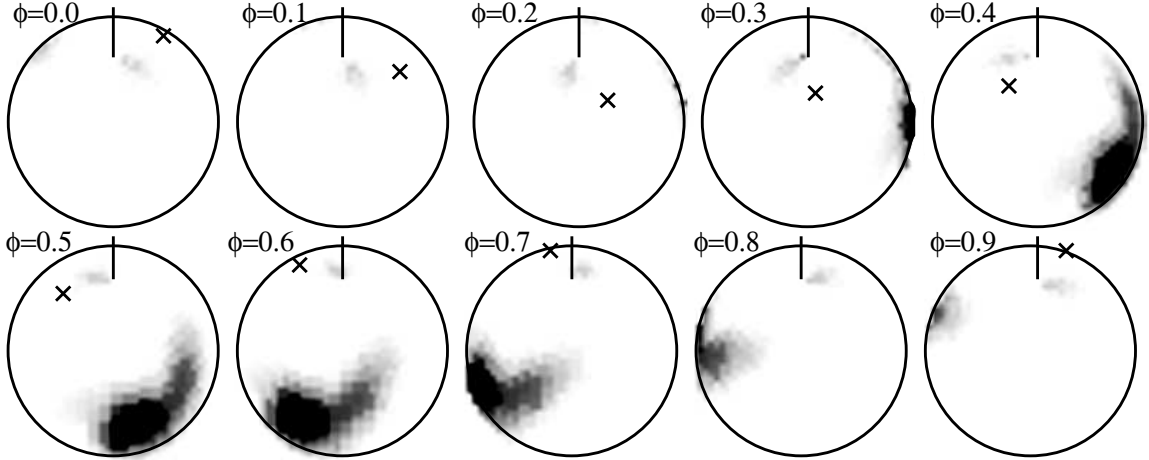}}
\end{picture}
\end{center}
\caption{The predicted location of the accretion regions on the white
dwarf as a function of spin phase for $i=42^{\circ}$,
and $\beta=50^{\circ}$. The magnetic axis is shown
as a cross.}
\label{globes} 
\end{figure*}

We show in Fig. \ref{globes} the position of the accretion regions
predicted by our model superimposed on globes representing a complete
rotation of the white dwarf for the same model fits as shown in
Fig. \ref{polfit}. An extended cyclotron accretion region, which is
offset by a large distance from the visible magnetic pole, is first
seen at $\phi\sim$0.3 (and may account for the increase in the linear
polarisation seen at this phase) and disappears at $\phi\sim$0.9 --
the primary accretion region. The `tail' of this region is optically
thin cyclotron radiation. Fig. \ref{globes} shows that the leading
edge of the primary region (optically thick cyclotron radiation)
quickly appears over the limb of the white dwarf and accounts for the
rapid increase in circularly polarised flux at $\phi\sim$0.3
(Fig. \ref{polfit}). On the other hand the `tail' of the accretion
region takes a longer time to disappear over the limb of the white
dwarf resulting in a less rapid decrease in the circularly polarised
flux.  A much smaller accretion region is predicted very close to the
spin pole and is always visible -- the secondary accretion region. For
higher values of $\beta$ ($>50^{\circ}$), we find other models which
fit the data equally well. In these cases the primary accretion region
is less extended in magnetic longitude and the secondary accretion is
brighter and vice-versa for $\beta <50^{\circ}$. We consider the
$\beta <50^{\circ}$ set of models to be the most likely solutions
since in these cases the secondary region is much less significant
than the primary and hence the most simple scenario (as is the
rationale of our optimisation technique).

While we are confident that all our best fitting models predict the
same general location, extent and phasing of the primary accretion
region, we cannot be certain of the finer details of its structure as
the resolution of the image is limited due to the finite sampling of
our data and the fact that the optimisation technique tends to smooth
out fine structure. We now go on to compare our model results with the
scenario of Misaki et al (1996) who suggest that the deep X-ray dip is
due to the accretion stream far from the white dwarf obscuring the
emission from the bright accretion region.

\section{Discussion}

The optical and X-ray data reported by Thomas \& Reinsch (1996) and
Misaki et al (1996) respectively are consistent with CP Tuc being a
polar although their data could not rule out an Intermediate Polar
interpretation (IPs - the non-synchronous magnetic CVs).  One of the
main characteristics of polars is their high levels of
polarisation. However, until now, no polarimetry data was available on
this object.  Our polarimetry shows circular polarisation of up to 10
per cent. Such high levels of polarisation together with the detection
of only one photometric period confirm that CP Tuc is a polar and not
an IP (IPs show complex amplitude spectra).

The one unusual feature that CP Tuc exhibits for a polar is the
prominent X-ray dip which lasts approximately half the orbital
cycle. Misaki et al (1996) suggested that this dip is caused by
obscuration of the accretion region by the accretion stream. This
conclusion was reached mainly due to the dip having a greater depth in
soft X-rays (1--4keV) compared to hard X-rays (4--13keV) which is
indicative of photo-electric absorption (Fig. \ref{folded}). In
contrast, the much broader dips seen in the EUV are grey. It is very
unusual for a polar to have an absorption dip lasting such a large
fraction of the binary orbit. In hard X-rays, absorption dips lasting
$\sim$0.1 of the binary orbit have been seen in a number of systems
(eg EF Eri: Done, Osborne \& Beardmore 1995), but dips lasting
$\sim$0.5 of the binary orbit have not.

Our modelling of our optical polarimetry data suggests that the bulk
of the optical flux originates from an extended accretion region far
from the visible magnetic pole. It is reasonable to assume that the
polarised optical flux and the X-ray flux originate close to the shock
region above the white dwarf. Thus, the dip seen in X-rays between
$\phi\sim$0.9--0.2 is simply due to the bulk of the primary accretion
region being self-eclipsed by the white dwarf according to our
polarisation modelling. However, even at the deepest point of the
X-ray dip a significant X-ray signal is present with hard X-rays being
more dominant than soft X-rays. 

By examining Fig. \ref{globes} it can be seen that the trailing edge
of the primary accretion region is still just visible at the deepest
point of the X-ray dip (although this maybe not be so clear in the
reproduction of Fig. \ref{globes}). In models where $\beta$ is
$>50^{\circ}$ this is more obvious. On the other hand the larger (and
leading) edge of the region is self-eclipsed for the phases centered
on the X-ray dip. If the leading edge of the region is predominately
emitting in soft X-rays and the trailing edge predominates in hard
X-rays then this can account for the energy dependance of the X-ray
dip.

The structure of the accretion region in polars depends on how and
where the accreting flow attaches to the magnetic field lines. A
magnetic field can be thought of as acting as a density filter in
which less dense material from the accretion flow is threaded first
while denser material is threaded futher along its trajectory. This
results in a variation in the local specific accretion rate over the
region and hence a range of X-ray energy. Our model suggests that (at
least in the case of CP Tuc) the trailing edge of the primary
accretion region has a higher mean X-ray temperature than the leading
edge.

Until this stage we do not know how our spin phase convention relates
to the binary orbital phase. We can use the optical spectroscopy of
Thomas \& Reinsch (1996) to roughly locate the position of the
secondary star. Thomas \& Reinsch obtained low resolution (32 \AA)
spectroscopy of CP Tuc and were able to resolve the H$\alpha$ line
into broad and narrow components. In low resolution spectra such as
these the broad component is likely to originate from the accretion
stream close to the white dwarf. On the other hand the narrow
component is likely to originate in the stream much further from the
white dwarf, perhaps closer to (or at) the secondary star.  The
maximum red shift of the broad and narrow components are expected to
occur just before inferior conjunction (the point where
the secondary is closest to the observer) and in that order.

The spectroscopic phase 0.0 of Thomas \& Reinsch (1996) corresponds to
our polarimetric phase 0.08.  From Fig. 3 of Thomas \& Reinsch (1996)
the broad component has a maximum blue shift at our $\phi\sim$0.5
while it reaches maximum red shift at our $\phi\sim$1.0. Examining
Fig. \ref{globes}, we can see that at $\phi\sim$0.4--0.5 the magnetic
field configuration is such the magnetic field lines feeding the
accretion region appearing over the limb of the white dwarf will
result in the accreting material being maximally blue shifted.
Similarly at $\phi\sim$0.8--0.9 our model predicts the accreting
material will have maximum red shift. Bearing in mind the
uncertainties in the relative phasing, our model results are
reasonably consistent with the spectroscopy of Thomas \& Reinsch
(1996).

Let us compare this result to the low resolution spectroscopy of the
eclipsing polar MN Hya (Ramsay \& Wheatley 1998) where $\phi$=0.0
defined the eclipse center. Ramsay \& Wheatley (1998) found a maximum
red shift for the broad component at $\phi\sim$0.9 and maximum blue
shift for the broad component at $\phi\sim$0.4. Therefore to obtain
the inferior conjunction of the secondary in CP Tuc we need to add
$\phi\sim$0.1 to the phase of maximum red shift, ie inferior
conjunction occurs at $\phi\sim$0.1. From the location of the
secondary star, it appears that the accretion flow has not accreted
onto the field lines of the more favourable magnetic pole (which is
never visible). Instead it has preferentially accreted onto the field
lines of the visible magnetic pole. This may suggest that the magnetic
field of the white dwarf maybe more complex than a dipole field as we
assume in the modelling of our polarisation data.

Finally, we briefly discuss the optical photometric observations which
show no modulation in $B$ or $V$, but show an increasingly large
modulation as we move further to the red (cf
Fig. \ref{folded}). Thomas \& Reinsch (1996) modeled their optical
spectra and determined a magnetic field strength $<$17MG. This is
consistent with the good fits we achieved in fitting our polarimetric
data with a fixed field strength of 15MG.  As the magnetic field
decreases, cyclotron radiation is shifted towards redder wavelengths
and hence for low field strengths light curves will be much less
modulated in the blue compared to the red, as is the case in CP Tuc.

\section{Acknowledgments}

We would to thank the Director of SAAO, Dr R Stobie, for the generous
allocation of observing time and Dr Darragh O'Donoghue for the use of
his period analysis software. We would like to thank the referee of
this paper, Dr H.-C. Thomas, for making some very helpful comments
which significantly improved this paper.

\end{document}